\documentclass[iop]{emulateapj}
\usepackage{color}
\usepackage{graphicx}
\usepackage{natbib}

\begin{document}

\title{X-ray Transients in the Advanced LIGO/Virgo Horizon}
\author{Jonah Kanner}
\affil{LIGO-California Institute of Technology, Pasadena, California 91125}
\affil{NASA Goddard Space Flight Center, Mail Code 663, Greenbelt, MD 20771}
\email{jonah.kanner@ligo.org}
\author{John Baker, Lindy Blackburn, Jordan Camp}
\affil{NASA Goddard Space Flight Center, Mail Code 663, Greenbelt, MD 20771}
\author{Kunal Mooley}
\affil{California Institute of Technology, Astronomy Department, Mail Code 249-17, Pasadena, CA 91125}
\author{Richard Mushotzky}
\affil{NASA Goddard Space Flight Center, Mail Code 661, Greenbelt, MD 20771}
\affil{Astronomy Department, University of Maryland, College Park, MD 20742}
\and
\author{Andy Ptak}
\affil{NASA Goddard Space Flight Center, Mail Code 661, Greenbelt, MD 20771}

\begin{abstract}
Advanced LIGO and Advanced Virgo will be all-sky monitors for merging 
compact objects within a few hundred Mpc.  Finding the electromagnetic 
counterparts to these events will require
an understanding of the transient sky at low red-shift (z $<$  0.1).  
We performed a systematic search for extragalactic, low red-shift, 
transient events in the XMM-Newton Slew Survey.
In a flux limited sample,
we found that highly-variable objects comprised 10\% of the sample,
and that of these, 10\% were spatially coincident with cataloged 
optical galaxies.
This led to $4 \times 10^{-4}$ transients per square degree 
above a flux threshold of $3 \times 10^{-12}$ erg cm$^{-2}$ s$^{-1}$ [0.2-2 keV] 
which might be confused with LIGO/Virgo counterparts.  
This represents the first extragalactic measurement of the soft X-ray transient 
rate within the Advanced LIGO/Virgo horizon.  
Our search revealed 
six objects that were spatially coincident with previously 
cataloged galaxies,
lacked evidence for optical AGNs, displayed high luminosities 
$\sim 10^{43}$ erg s$^{-1}$, and varied in flux by more than a 
factor of ten when compared with the ROSAT All-Sky Survey.
At least four of these displayed properties consistent with
previously observed tidal disruption events.
\end{abstract}


\maketitle

\section{Introduction}
The X-ray band provides an opportunity to find high 
confidence counterparts to the compact
object mergers that will be discovered with the second generation 
LIGO and Virgo gravitational wave
detectors. Within this decade Advanced LIGO \citep{advLigo} 
and Advanced Virgo\footnote{https://tds.ego-gw.it/itf/tds/file.php?callFile=VIR-0027A-09.pdf} \citep{advVirgo}
are expected to begin detecting mergers of
binary neutron stars and neutron stars with stellar mass black 
holes out to distances of a few hundred Mpc \citep{cbcRates}.
Placing these mergers in an astrophysical context and 
maximizing the scientific returns will require 
finding electromagnetic counterparts to the
events \citep{bloomDec, phinney09}. 
However, the positional accuracy of the gravitational
wave detectors will be limited to 
tens or hundreds of square degrees
\citep{methods, fairhurst, cwbposrec, nissankee2e}. 
Thus, associating
an electromagnetic counterpart with a LIGO/Virgo 
detection will require an understanding that a chance
coincidence within the LIGO/Virgo horizon is unlikely, 
even within a large sky region.  In the optical band,
large area survey instruments
such as Pan-STARRS, Palomar Transient Factory,
SkyMapper, and the future LSST 
will have a daunting challenge separating LIGO counterparts
from stellar variability, supernovae, and other confusion sources \citep{kulkarni09, nissankee2e},
but may leverage the large on-going effort to create
schemes of automated or semi-automated transient
classification (e.g. \citet{ptf_tcp}).
The radio band may also be searched for counterparts to GW events
\citep{predoi, lazio2012}, and large area searches for
radio transients are rapidly developing \citep{lofar2011, bhat2013}.

In the X-ray band,
low number counts at bright flux levels 
may make identification of a LIGO/Virgo counterpart
more straightforward.  However, studies of 
X-ray variability have tended to focus on 
persistent or repeating sources, particularly 
AGN, X-ray binaries, or stellar flares.   
Past all-sky searches for soft X-ray transients 
include searches in the 
ROSAT All-Sky Survey (RASS) data \citep{rass_grb, rass_var},
work with the XMM-Newton 
Slew Survey \citep{esquej07, starling11}, and searches for 
flashes lasting a few seconds \citep{flash_1, flash_2}.

A small fraction of compact object mergers are thought
to create short gamma-ray bursts (GRBs) \citep[e.g.][]{fox05, eichler89},
and if one occurred within the Advanced LIGO/Virgo horizon,
the X-ray afterglow could be bright ($\sim 10^{-12}$ erg cm$^{-2}$ s$^{-1}$)
for about a day after the merger \citep{kanner12}.
Short GRBs may be associated with mergers of two neutron stars, 
or a neutron star with a black hole, and are typically characterized
by a prompt emission lasting less than two seconds, and a spectrum
that is somewhat harder than the more prevalent ``long GRBs'' associated with
stellar core-collapse \citep{nakar07}.  
Some double neutron star mergers may also exhibit a 
bright X-ray counterpart due to
emission from a magnetar-powered ejecta \citep{gao, zhang, metz_mag},
observable for up to a few thousand seconds after the merger.
Such signals could be detectable with a wide field 
focusing instrument such as the proposed ISS-Lobster 
\citep{camplobster} or A-STAR \citep{astar}, 
or in some cases with multiple
observations of the Swift XRT \citep{kanner12, swift_lvc}.
In order to estimate the ability to identify a unique source
in such a campaign,
we have searched
the XMM-Newton Slew Survey Clean Source Catalog
version 1.5 (XMMSL1, \citet{xmm_slew}) for objects consistent
with counterparts to compact object mergers
observable with the future Advanced LIGO and 
Advanced Virgo observatories.  
We sought both to measure the density of 
such events that are observable in a given moment, 
and to characterize the nature
of the objects we found.  
Our search criteria emphasized low red-shift, transient objects, and
were similar to those of \citet{esquej07}.  However, this study 
used a data set covering five times the slew survey area available in 2007 and 
included a systematic measurement of the transient density using
a simple, easy to emulate definition.

\section{Candidate Selection} \label{selection}

\begin{figure}
\begin{center}
\mbox{
\includegraphics*[width=0.50\textwidth]{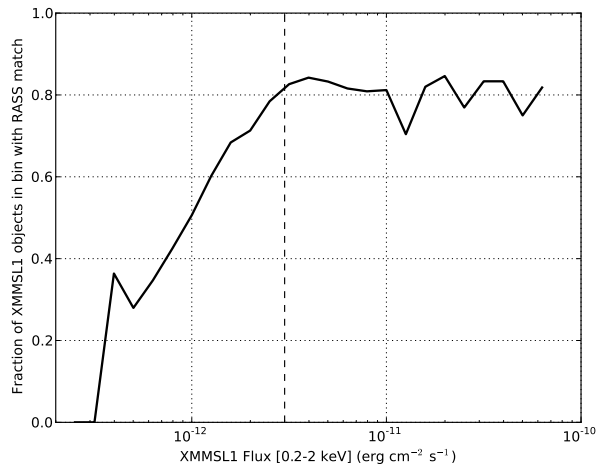} 
}
\caption{The fraction of XMMSL1 objects listed with RASS matches in each 
flux bin.  The vertical line indicates the flux limit selected for the search.}
\label{frac_match}
\end{center}
\end{figure}

\begin{figure}
\begin{center}
\mbox{
\includegraphics*[width=0.5\textwidth]{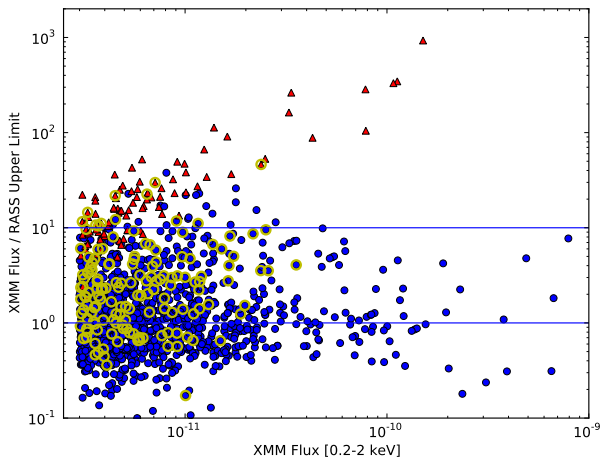} 
}
\caption{Sources above our flux threshold in the XMMSL1.  The y-axis
represents the flux ratio between the XMMSL1 and RASS.  All RASS fluxes are calculated as 2-sigma upper limits, with red triangles indicating a number of RASS counts consistent with a non-detection at the 2-sigma level, and blue circles indicating a source was detected.  A yellow, hollow circle indicates an identification with a galaxy in the XMMSL1.  }
\label{scatter}
\end{center}
\end{figure}

We designed our selection criteria to seek objects
consistent with a transient event within the 
Advanced LIGO/Virgo neutron star merger horizon distance.  We used the XMMSL1
soft band of 0.2-2 keV for all flux measurements, since
it is similar to the ROSAT PSPC band.
The XMMSL1 includes identifications with RASS sources with
a 30'' search radius \citep{rass_bsc, xmm_slew}.  We found that 80\% of 
XMMSL1 sources brighter than $3 \times 10^{-12}$ erg s$^{-1}$ cm$^{-2}$ had matches in the 
RASS with no improvement
in overlap for brighter sources, so we took this as the 
flux limit for our search (See Figure \ref{frac_match}).  
For each XMMSL1 object brighter than this threshold, we attempted to
place a flux upper-limit in the RASS data set.  
We explored a range of different radii for the source extraction region 
using data corresponding to ROSAT detected sources, 
and found that a 205 arcsecond
radius was 
needed to recover the median source with 90\% of the 
expected flux.
For the XMMSL1 objects above our flux threshold, 
Bayesian upper-limits corresponding
to a 2-sigma confidence level were applied to the ROSAT 
PSPC counts found within a 205 arcsecond extraction radius.  
To convert from counts to flux in the 0.2-2 keV band, we used 
webPIMMS\footnote{http://heasarc.gsfc.nasa.gov/Tools/w3pimms.html}
with 
the same source assumptions that were used in the 
XMMSL1, namely, a power-law spectra with an index of 1.7 and
a Hydrogen column density of $3 \times 10^{20}$ cm$^2$.
For each object, we took a ratio between XMMSL1 flux as listed in the 
catalog and
the RASS upper-limit we constructed, and kept only
objects with a flux ratio greater than ten (See Figure \ref{scatter}). 
These objects, or any objects with a flux ratio greater than ten, are referred
to as ``transients'' for the rest of this paper.
Since we were interested
only in extragalactic objects within a few hundred Mpc, 
we expected their host galaxies
to be visible in large optical and near infrared surveys such as 
the Sloan Digital Sky Survey\footnote{http://www.sdss.org} (SDSS) and the Two Micron All Sky Survey (2MASS).  
So, we demanded
that candidates be listed in XMMSL1 as spatially coincident with 
a known galaxy or galaxy cluster with a 30'' match radius.  
Finally, we checked our
candidates for an AGN association, and for any other observations
in the HEASARC database\footnote{http://heasarc.gsfc.nasa.gov/}.

\section{Results}

\begin{figure}
\begin{center}
\mbox{
\includegraphics*[width=0.5\textwidth]{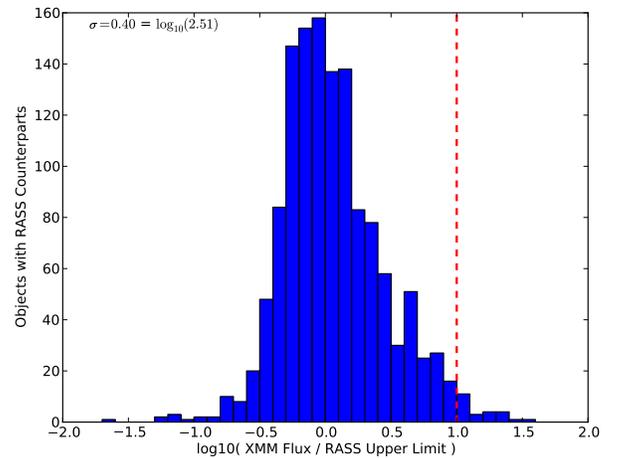} 
}
\caption{Histogram of the logarithm of the flux ratio for 
objects detected in both XMMSL1 and RASS.  The standard deviation
of the distribution is 0.40, corresponding to a flux ratio of 2.51.
By this measure, then, a threshold on the flux ratio of 10 requires an
object's variability to be beyond 2.5 $\sigma$ in the distribution.  }
\label{histogram}
\end{center}
\end{figure}

\begin{figure}
\begin{center}
\mbox{
\includegraphics*[width=0.5\textwidth]{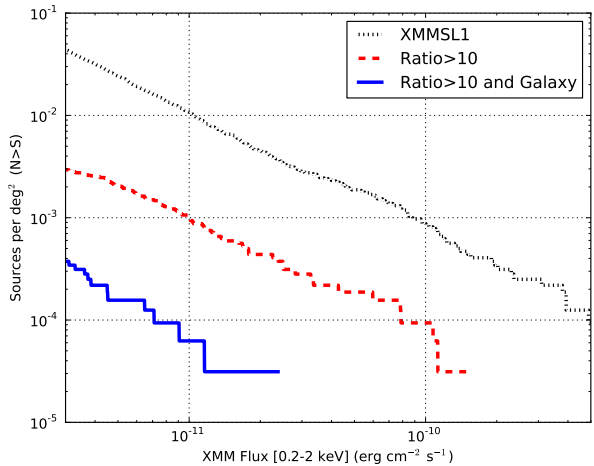} 
}
\caption{The statistical result of our search for low-redshift transients.
The top curve (black) shows the logN-logS plot for the XMMSL1 catalog,
containing 1411 objects above our flux threshold.  The 2nd curve (red) shows 
the distribution of the 97 transients, defined as at least ten times brighter in XMMSL1 than in RASS.  
The bottom curve includes only transient objects that are  
spatially coincident with a known galaxy, 
after rejecting previously identified AGN.}
\label{tran_count}
\end{center}
\end{figure}

\begin{table*}
\begin{center}
  \begin{tabular}{ l l l l l l l}
    \hline
	Name     & Flux & Ratio     & $D_L$\tablenotemark{a} & Luminosity & Notes & Date \\
    XMMSL1\dots& (cgs) &          & (Mpc)    & (erg s$^{-1}$)    &       &    \\       
    \hline
    J202320.7-670021 & $2.3 \times 10^{-11}$ & 46 & 67   & $1.2\times10^{43}$  & Not AGN\tablenotemark{c} & 2009-10-05 \\ 
    J084837.9+193527 & $6.4 \times 10^{-12}$ & 23 & 283  & $6.1\times10^{43}$  & Emission Line Galaxy\tablenotemark{d} & 2010-10-17 \\
    J182609.9+545005 & $4.5 \times 10^{-12}$ & 22 & 650  & $2.2\times10^{44}$  & Not AGN                  & 2005-11-01 \\
    J152408.6+705533 & $3.3 \times 10^{-12}$ & 15 & 256  & $2.6\times10^{43}$  & Spiral Galaxy & 2006-01-24, 2007-11-01 \\
    J202554.8-511629 & $9.1 \times 10^{-12}$ & 12 & ?    & ?                  &                         & 2010-04-16 \\
    J131951.9+225957 & $3.7 \times 10^{-12}$ & 10 & 99   & $4.3\times10^{42}$  & Not AGN\tablenotemark{d} & 2005-07-15 \\
    \hline
    J111527.3+180638 & $7.1 \times 10^{-12}$ & 30 & 12   & $1.2\times10^{41}$ & Esquej.        & 2003-11-22 \\
    J155631.5+632540 & $3.8 \times 10^{-12}$ & 21 & ?    & ?             & Bad match      & 2010-06-23 \\
    J020303.1-074154 & $3.1 \times 10^{-12}$ & 12 & 272  & $2.7\times10^{43}$ & Esquej.         & 2004-01-14 \\
    J170543.0+850523 & $4.5 \times 10^{-12}$ & 12 & ?    & ?             & Cluster        & 2008-08-31 \\
    J013727.9-195605 & $1.2 \times 10^{-11}$ & 11 & 1320 & $2.1\times10^{45}$ & Cluster         & 2007-12-27  \\
    J104745.6-375932 & $3.6 \times 10^{-12}$ & 10 & 335  & $4.8\times10^{43}$  & AGN\tablenotemark{c}  & 2003-12-14  \\
    \hline
\end{tabular}
\label{table1}
\caption{\footnotesize
The list of transient objects with galaxy associations found in our search.  The listed fluxes are those 
reported in XMMSL1 catalog \citep{xmm_slew}.  The column labeled
``Ratio'' shows the ratio between the observed XMMSL1 flux and a 2-sigma
upper limit based on the corresponding RASS data.  Each distance (D$_L$) and 
inferred luminosity is based on the overlap of the 
2-sigma XMMSL1 position with an optically identified galaxy.  
The note ``Not AGN'' signals that an available spectra 
shows no evidence for AGN emission.  The objects
with the note Esquej were reported as tidal disruption event candidates
by Esquej et al. (2007).  Bad match denotes that the matched
galaxy was separated from the XMMSL1 source by more than the 2-$\sigma$ position uncertainty.
The six objects above the horizontal line are further discussed in Section 3.
}
\tablenotetext{1}{The luminosity distance of the associated galaxy}
\tablenotetext{2}{Based on inspection of 6dF spectrum}
\tablenotetext{3}{Based on inspection of SDSS spectrum}
\end{center}
\end{table*}  

\begin{table}
\begin{center}
	\begin{tabular}{lll}
	\hline
	Name & Galaxy match & offset\\
	XMMSL1\dots & & [asec] \\
	\hline
		J202320.7-670021 & 6dFGS gJ202322.7-670046 & 27 $\pm$ 13 \\
		J084837.9+193527 & SDSS J084838.57+193528.9 & 10 $\pm$ 22 \\
                J182609.9+545005 & 2MASX J18261094+5450052 & 7 $\pm$ 5 \\
		J152408.6+705533 & MRK 1097 & 1 $\pm$ 3 \\		
                J202554.8-511629 & 2MASX J20255579-5116276 & 9 $\pm$ 19     \\
                J131951.9+225957 & NGC 5092 & 4 $\pm$ 2 \\
	\hline
\end{tabular}
\label{table2}
\caption{\footnotesize
Galaxy matches from the XMMSL1 catalog for six transients identified by our search.  The offset shows the angular distance between the galaxy and the source position, along with the uncertainty quoted in the catalog.}
\end{center}
\end{table}

\subsection{A snapshot of the sky}

Our search showed that source variability
seemed to naturally divide the XMMSL1 survey into two classes (See Figure \ref{scatter}).  
The first class, representing
roughly 90\% of sources, was mainly below the flux ratio threshold value of
ten, and mainly detected in both the RASS and the XMMSL1, indicated by 
blue circles in the figure.  The distribution of sources detected in both
XMMSL1 and the RASS is shown in Figure \ref{histogram}, and is
loosely consistent with a lognormal distribution. 
Their flux ratios
have a logarithmic standard deviation corresponding to a 
flux ratio of 2.5, and a flux ratio of more than ten corresponds to a
2.5-$\sigma$ level of variability.  
We label this class ``continuum variability'' and note the variability
arises from a wide range of causes, including
most AGN variability, measurement errors, and differences in the spectral 
response
of the two instruments.  

A different, more dramatic, type of variability
was also present in the survey.  At high flux values 
(towards the right in Figure \ref{scatter}), there
is a clear separation between the sources which were observed in the RASS, 
and those which were not observed in the RASS.  The ratios between the 
XMMSL1 flux and the RASS upper-limit for sources not detected in RASS are
plotted as red triangles in the figure.
This second population (state-change objects)
represents sources which were in some distinctly different state at the 
time of the RASS and XMMSL1 observations.  Known examples in this class
include X-ray binaries in different states, flaring stars, 
tidal disruption events, and classical novae.  
This class could be defined as sources 
with a flux and flux ratio which places them outside the distribution 
of continuum variability shown in in Figure \ref{histogram}.  
Many of the sources with RASS non-detections
(red triangles) seem to belong to this class.  Though the separation between 
the RASS detections and non-detections in Figure \ref{scatter} disappears
at low flux values, the vertical placement of the non-detections is 
determined primarily by the limiting flux of the RASS observation.  So,
any of the non-detections would potentially sit higher in this plane 
with data from a deeper observation.  Given that some of the bright XMMSL1
sources were seen to vary by more than a factor of 100, 
it would be surprising if this was not also the case for some of the 
dimmer sources as well.  

Some of the state-change objects observed in XMMSL1 were also 
studied by \citet{starling11}.
In their study, the authors selected 97 sources from v1.4
of the XMM Slew Survey with no identified optical or RASS counterpart.
They then collected pointed observations for 94 of these with the \textit{Swift} 
X-ray Telescope in an attempt to classify them.
In their paper, \citet{starling11} report that 
71\% of the targets were not seen in the \textit{Swift} observations, implying
they had faded by at least a factor of ten.   
The isotropic distribution and lack of 
an optical counterpart of these mysterious sources led the authors to conclude 
that they represented a primarily extragalactic population.  
None of the sources in the Starling study appeared in our final candidate list,
as the former campaign required sources to have no optical counterpart, 
where we required a match to a known optical galaxy.  However, if the 
bulk of the transients studied by Starling et al. were extragalactic, 
as they concluded, then the objects listed in Table 1 may represent
low red-shift objects from the same population.  In particular, the study
found 47 objects detected in the XMM-Newton soft-band, but not
detected with \textit{Swift}.

We wished to use the variability in this data set to characterize the 
chance of a spurious detection in coincidence with a trigger from the 
future Advanced LIGO/Virgo network.  We applied the selection 
criteria described in Section \ref{selection}, and the 
statistical results of our search are presented in Figure
\ref{tran_count}.
Each curve in the figure shows the density of sources in the set
with fluxes above the value on the X-axis.  The top curve (black, dotted) is 
the logN-logS curve of the XMMSL1
catalog, assuming 32,800 square degrees of coverage.  The 
bottom curve (blue, solid) is the result of our search criteria, which
resulted in 12 transient objects spatially coincident with a known 
galaxy (See Table 1, yellow circles in Figure \ref{scatter}).
In requiring the galaxy association, we included associations with
both galaxies and galaxy clusters, but rejected associations
labeled in the XMMSL1 as AGN (Seyfert, BL Lac, etc.).  

Most of the matched host galaxies with available redshifts were
located at luminosity distances of less than 350 Mpc.  This is 
most likely because available catalogs of galaxies were dominated
by relatively shallow observations from large area surveys.  
\citet{mansi_phd} found that current galaxy 
databases (NED\footnote{http://ned.ipac.caltech.edu/}, 
Hyperleda\footnote{http://leda.univ-lyon1.fr/}, etc.) were 50\% 
complete to 200 Mpc,
with a steep downward trend in completeness as a function of
distance.  
This distance scale was well aligned with 
the reach of Advanced LIGO and Virgo, which will detect 
NS-NS mergers primarily between 100 and 400 Mpc \citep{nissankee2e}.
From an operational perspective, then, future
searches for X-ray counterparts to LIGO/Virgo transients may not need
to obtain a red-shift estimate for each possible host galaxy
in the field; any galaxy bright enough to be cataloged by a large area 
survey without an AGN signature is likely to be within the Advanced LIGO
horizon.  For our interpretation, we have taken this approach.  This 
would tend to skew our estimates of unrelated counterparts to be artificially
high, because using the estimated distance to the gravitational wave 
source can be a powerful tool for rejecting potential host galaxies
at inconsistent redshifts \citep{nissankee2e}.

To interpret Figure \ref{tran_count}, we note that typical exposures
in the XMM Slew Survey were 5-20 seconds, so we assumed that the 
transients that passed our selection criteria were of longer duration
than the XMM-Newton exposures.  Though the Slew Survey was not 
uniform in any sense, the coverage area was largely random, and so
we assumed that the density of sources which passed our cuts was not strongly
biased by target selection.  The survey did include some repeat visits, 
and only covered 20,900 unique square degrees. However,
time between repeated visits was typically $> 1$ year, 
which was more than 
the expected time-scale for fading of X-ray counterparts to neutron 
star mergers.  For this reason, we interpreted our results based
on 32,800 square degrees of coverage, noting that this choice 
leads to at most a 60\% systematic in our results.  
So, for example, the 12 sources above our flux threshold may be interpreted as 
$4 \times 10^{-4}$ transients per square degree on time-scales
shorter than the $\sim$15 years between RASS and XMMSL1.

The positional uncertainties associated with a trigger from the 
LIGO/Virgo network are expected to vary a great deal depending
on signal-to-noise ratio and other factors, however, studies typically
quote numbers between 20 and 200 square degrees for the position
uncertainty of a low signal-to-noise ratio trigger with three
sensitive gravitational wave detectors operational \citep{methods, fairhurst, cwbposrec}.  Searching 
such a large area for an X-ray counterpart will require a 
relatively high flux limit
to the search, since the instrument will have to either be 
very wide field or will have to use short exposures for many tiles.  
For example, in principle the 
Swift XRT could utilize $\sim 100$ s exposures to tile as much as 
35 deg$^2$ in a day, though only to a depth of 
$6 \times 10^{-12}$ erg cm$^{-2}$ s$^{-1}$ \citep{kanner12}. While practical issues concerned with 
many repointings of the instrument may make this difficult
in practice, Figure \ref{tran_count} shows that there would be less
than a 1\% chance of finding a transient in this search area 
by chance.  A more natural scenario would be the application of a 
very wide-field, focusing instrument such as the proposed ISS-Lobster.
The proposed instrument would image a 400 deg$^2$ field of view to 
a depth of around 10$^{-11}$erg s$^{-1}$cm$^{-2}$ in a twenty minute exposure.  
Applying these numbers, this observation 
would have a 3\% chance of imaging an unrelated transient coincident 
with a known host galaxy, or less than a
1\% chance if we imagine only using the fraction of the field that overlaps
the LIGO/Virgo errorbox.  Similar considerations would apply to
other wide field-of-view instruments, including MAXI \citep{maxi} or the proposed
A-STAR \citep{astar}.  

Of course, the details of how a future search is carried out would have a 
strong influence on these numbers.  Choosing to define a transient
as an object that brightens by at least a factor of ten is somewhat arbitrary,
though for our data set, the naturally distinct classes of 
continuum variability and state-change variability seen in Figure 
\ref{scatter} appears to somewhat justify this choice.  Another important 
factor is the completeness of the galaxy catalog, which is likely to 
evolve rapidly due to efforts by several large area surveys \citep{metzger_catalog}.  
On the other hand, at the order-of-magnitude level, these results seem
to be robust.  The associations of these sources with the host galaxies
do not appear to be spurious (See section \ref{spurious}), so adjusting the 
match radius used to associate host galaxies and sources will make little
impact.  Similarly, increasing the searched area around the host galaxy
due to ``kicks'' in the binary \citep{kalogera_kicks} will not change these statistics
substantially.  A variety of models, with 
some validation from studies of short GRB host galaxies,
have concluded that the majority of NS/NS mergers occur within ten or a 
few tens of kpcs from the centers of their host galaxies \citep{berger_nohost, brandt_kicks, bloom_kicks, fryer_99, belczynski_kicks}.  At 
200 Mpc, the 30 arcsecond search radius used for galaxy association 
corresponds to an offset 
from the host galaxy of 30 kpc.  \citet{berger_nohost} showed that
for a range of models, this search radius would include 70-90\%
of NS/NS mergers.  On the other hand, there do exist models with 
more extreme kick velocities,
leading to mergers that occur up to a Mpc away from the host galaxy
\citep{kelley_kicks}.  To 
accommodate these models would mean using a somewhat larger search radius, and 
so the requirement of a galaxy association may become less useful.  
Additional surveys could increase the number of known galaxies, and so increase 
false associations by a factor of $\sim 2$ or more. 
However, eliminating possible hosts with redshifts inconsistent 
with the GW data could limit this effect\citep{nissankee2e}.  

Finally, we note that the separation of soft X-ray sources into state-change 
transients and continuum variability appears to occur naturally.  
For these reasons, we expect that the finding that around 10\% of 
bright, soft X-ray sources demonstrate a state-change brightening over long 
time scales, and that around 10\% of these can be associated with 
galaxies within 200 Mpc, will prove true for future searches with perhaps 
a factor of a few
uncertainty in both cases.  
For
searches for X-ray counterparts to Advanced LIGO/Virgo triggers, 
this criteria represents a two order of magnitude reduction in the background
rate, as compared with the density of X-ray sources on the sky which has
been used to estimate the density of spurious associations in past
work \citep{swift_lvc}.

\subsection{Further selection}

\begin{figure}
\begin{center}
\mbox{
\includegraphics*[width=0.5\textwidth]{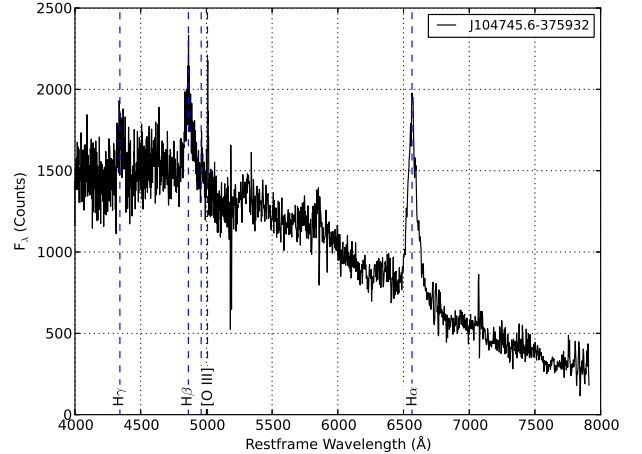} 
}
\caption{Optical spectrum of the host galaxy matched to XMMSL1 J104745.6-375932 obtained by the 6dF survey, plotted in the restframe using a redshift of 0.075.  The strong, broad Hydrogen lines and the blue spectrum are characteristic of quasars.}  
\label{spec_J1047}
\end{center}
\end{figure}

\begin{figure}
\begin{center}
\mbox{
\includegraphics*[width=0.5\textwidth]{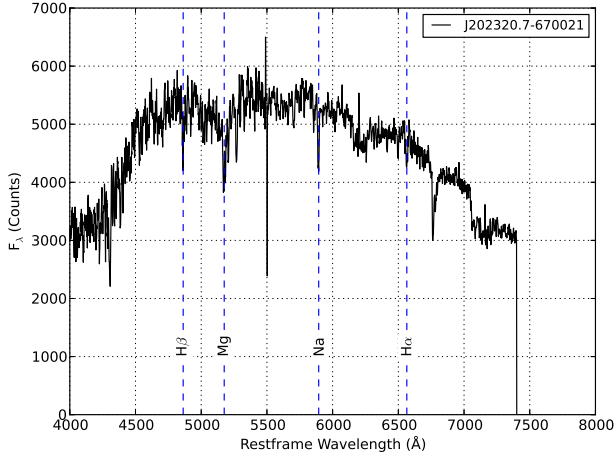} 
}
\caption{Optical spectrum of the host galaxy matched to XMMSL1 J202320.7-670021, obtained by the 6dF survey, and plotted in the restframe using
a redshift of z=0.016.  The galaxy is dominated by stellar absorption 
features.  }  
\label{spec_J2023}
\end{center}
\end{figure}

\begin{figure}
\begin{center}
\mbox{
\includegraphics*[width=0.5\textwidth]{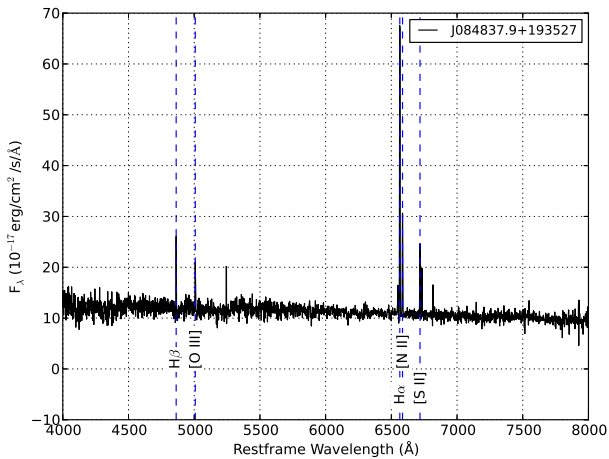} 
}
\caption{Optical spectrum of the host galaxy matched to XMMSL1 J084837.9+193527, obtained by the SDSS, plotted in the restframe using a redshift of 0.064.  A test using diagnostic narrow line ratios shows this to be a star forming galaxy, as described in the text.}  
\label{spec_J0848}
\end{center}
\end{figure}

\begin{figure}
\begin{center}
\mbox{
\includegraphics*[width=0.5\textwidth]{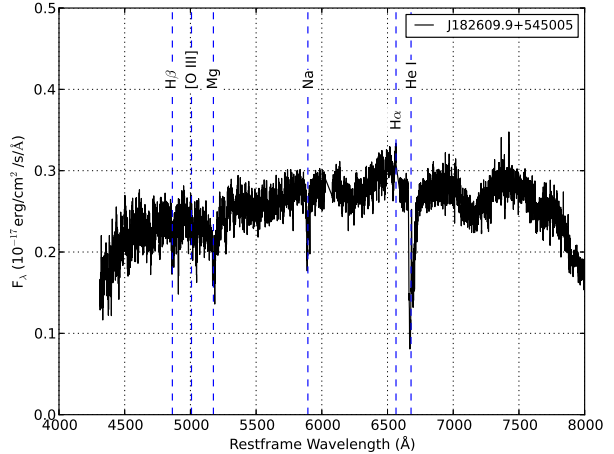} 
}
\caption{Optical spectrum of the host galaxy matched to XMMSL1 J182609.9+545005, 
obtained with the Keck DEIMOS on May 3, 2013,
plotted in the rest frame with a redshift of 0.14.  The absence of emission
lines and red color suggest this is a red galaxy without an active central region.}
\label{spec_J1826}
\end{center}
\end{figure}

\begin{figure}
\begin{center}
\mbox{
\includegraphics*[width=0.5\textwidth]{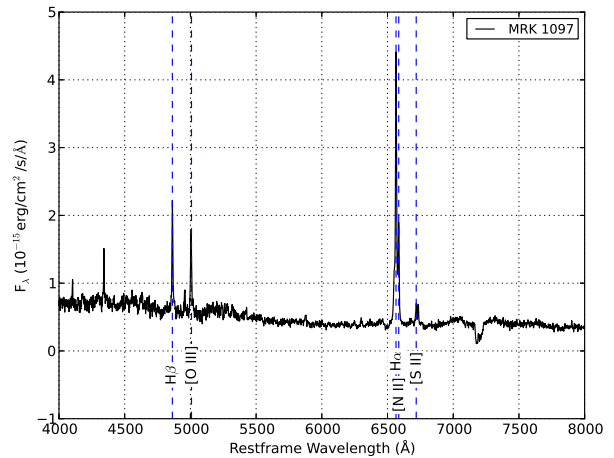} 
}
\caption{Optical spectrum of the host galaxy matched to XMMSL1 J152408.6+705533, obtained on April 15, 2013 with the 200-inch Hale Telescope at Palomar Observatory, 
plotted in the rest frame with a redshift of 0.059.  The narrow line features
and WISE colors suggest this is a spiral galaxy, and so is unlikely
to host an active nucleus.}
\label{spec_J1524}
\end{center}
\end{figure}

\begin{figure}
\begin{center}
\mbox{
\includegraphics*[width=0.5\textwidth]{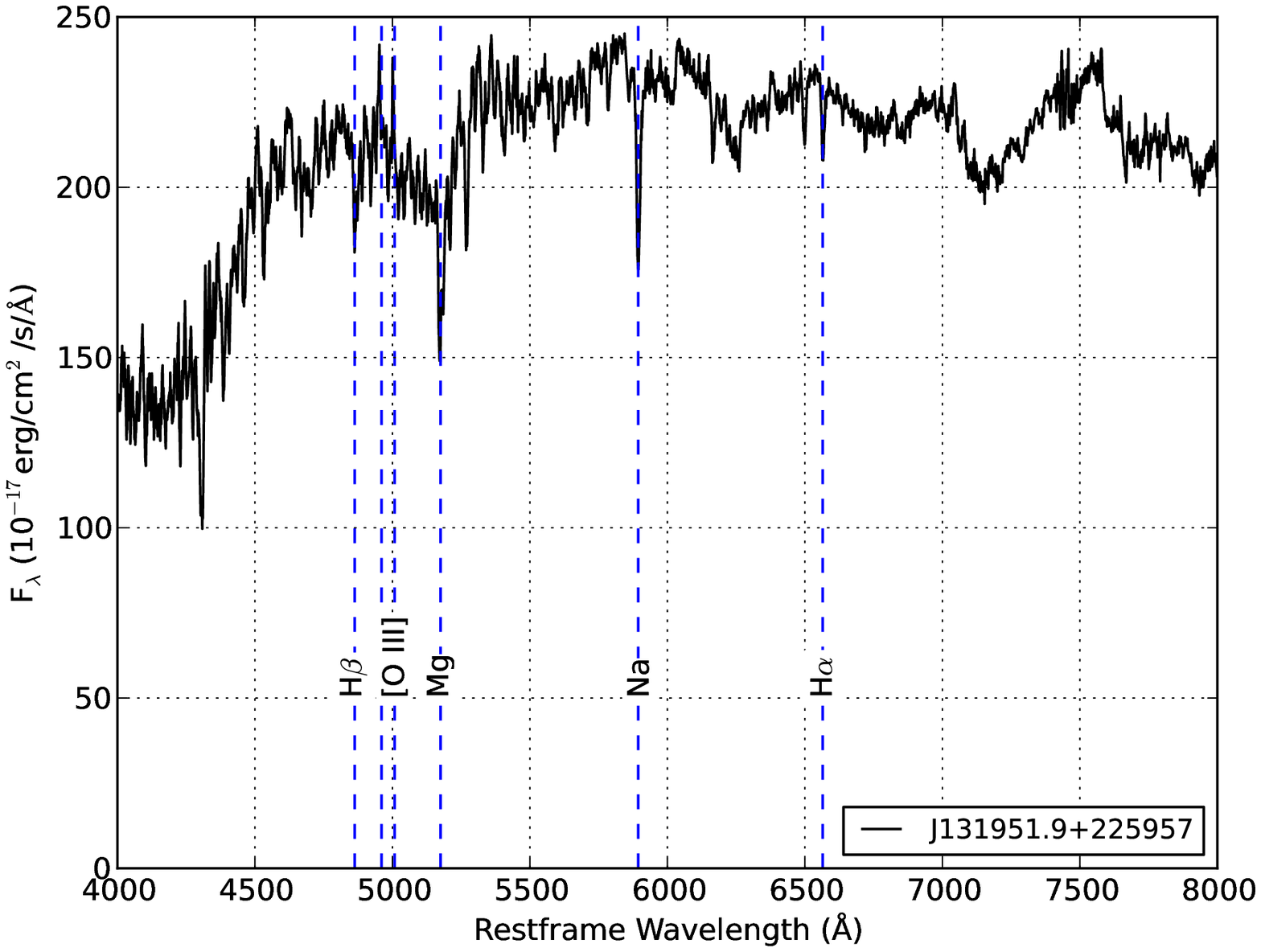} 
}
\caption{Optical spectrum of the host galaxy matched to XMMSL1 J131951.9+225957, obtained by the SDSS, plotted in the rest frame with a redshift of 0.023.  The spectrum appears to be dominated by stellar absorption features.}  
\label{spec_J1319}
\end{center}
\end{figure}

After identifying our list of twelve candidates, we sought
to characterize them as far as possible.  We inspected optical 
images of the host galaxies and searched for additional data using the
HEASARC database.  We also inspected publicly available and newly 
obtained optical spectra
to check for AGN signatures.  Two of the objects
were rediscoveries of previously published candidate tidal disruption
events (XMMSL1 J020303.1-074154 and XMMSL1 J111527.3+180638) \citep{esquej07}.  
One object showed broad AGN emission lines
in a 6dF\footnote{http://www.aao.gov.au/6dFGS} archived spectrum, and was seen in XRT data to have
an X-ray spectrum consistent with an AGN (XMMSL1 J104745.6-375932; See Figure \ref{spec_J1047}).  
One object's (XMMSL1 J155631.5+632540)
2-sigma position circle did not include the 
matched host galaxy.  Two of the sources 
(XMMSL1 J013727.9-195605 and XMMSL1 J170543.0+850523)
were matched 
to galaxy clusters with ambiguous galaxy associations.

This left six objects which exhibited some interesting properties.
Of the five where we obtained a redshift (either through observations or the literature),
most corresponded to luminosity distances less than 300 Mpc in a 
standard cosmology; the furthest was placed around 650 Mpc.
We examined optical spectra for five of the host galaxies,
and none of these showed broad
emission lines, though two (XMMSL1 J084837.9+193527 and J152408.6+705533) showed
narrow emission lines (See Figures \ref{spec_J2023} -- \ref{spec_J1319}).  
They were all soft sources, and only one has a measured hardness ratio
presented in XMMSL1 (XMMSL1 J182609.9+545005: -0.46).
We searched HEASARC for observations
with  
\textit{Chandra, Swift}, and 
\textit{XMM-Newton}, but found no observations
containing any of the six remaining sources.
These six objects were generally
extragalactic, more powerful than 4$\times 10^{42}$ erg s$^{-1}$, 
close ($D_L < 300$ Mpc), 
variable by at least 10$\times$, and lacking evidence for AGN activity.  
This makes them difficult to characterize, as well as energetic and
variable.  In the next section, we discuss possible characterizations
for these sources, and attempt to address the likelihood of each.

\section{Possible Characterizations}

\subsection{Tidal Disruption Events}

Tidal disruption events (TDE) are thought to occur at a rate of
$\sim 10^{-5}$ yr$^{-1}$Mpc$^{-3}$ and display initial luminosities
up to $\sim 10^{45}$ erg s$^{-1}$ during the first few days or weeks 
followed by a characteristic dimming $\propto t^{-5/3}$ \citep{tde_1, tde_2}.  
Under these assumptions, at 200 Mpc, such an event would be 
visible for a few months to years above a flux of $3 \times 10^{-12}$
erg s$^{-1}$ cm$^{-2}$.  
A number of TDE candidates have been discovered through their X-ray 
emission 
\citep{komossa99, grupe99, komossa_greiner99, greiner2000, maksym2010, lin2011}.
In addition, there have been discoveries made using ultraviolet surveys, such as those 
discussed in \citet{gezari2006, gezari2012}, and with optical surveys \citep{sjoert2011, drake2011}.  There have also been two well-studied TDEs discovered
with the \textit{Swift} satellite \citep[e.g.][]{bloom2011_swiftTDE, levan_swiftTDE, burrows_swiftTDE, cenko2012_swiftTDE}.

In fact, some TDEs have already been discovered using data
from the XMM-Newton Slew Survey.  Last year, \citet{saxton2012} reported on 
the discovery of an X-ray flare in SDSS J120136.02+300305.5 found in 
data from the XMM Slew Survey, and followed up with pointed observations
of the \textit{Swift} X-ray Telescope.  
Their program is designed to find flaring X-ray events soon after they begin
to enable prompt follow-up.
The flux of the corresponding observation
listed in the XMMSL1 was below our flux threshold, so this 
event was not rediscovered by our search.
\citet{esquej07, esquej08} found two TDEs in the XMM-Newton Slew Survey,
including one TDE above our flux threshold (See Table 1), via 
a search with similar selection criteria, but using
the first release (v1.1) of the XMMSL1 catalog, covering 6300 
square degrees.  Our search used the latest release (v1.5) of XMMSL1,
covering 32,800 square degrees (20,900 square degrees of unique area).
This suggests that we should expect
around four new TDEs in our data set.  Given the lack of nuclear
activity in some of the host galaxies, the high luminosity of the sources, 
and the positions which are consistent with the center of the host galaxies,
we expect this model will account for at least some of our candidates. 
For example, XMMSL1 J131951.9+225957 was matched to NGC 5092 with an 
offset from the center of the galaxy of only 4 arcseconds, compared with
the match radius of thirty arcseconds, or the average XMMSL1 uncertainty
of 8 arcseconds (See Table \ref{table2}).  An SDSS spectrum
of this galaxy, shown in Figure \ref{spec_J1319}, reveals only absorption features, making an AGN association unlikely.  These features were consistent with 
a TDE description. 
The galaxy SDSS J084838.57+193528.9 showed some narrow emission line 
features.  To characterize it, we used the best fit line profiles provided
by the SDSS Science Archive Server.  We found the 
[NII]$\lambda6583$/H$\alpha$ ratio to be 0.32, and the 
[OIII]$\lambda5007$/H$\beta$ ratio to be 0.67, so that this galaxy
appeared to have a star-forming region, but no nuclear activity
\citep{veilleux}.  For this 
reason, the observed variability in this galaxy also appeared consistent 
with a tidal disruption model.
We obtained a spectrum of the host galaxy matched to 
XMMSL1 J182609.9+545005 on May 3, 2013, using the Keck DEIMOS
(Figure \ref{spec_J1826}).  The spectrum was dominated by absorption features,
and showed no evidence for an active central region.  The red-shift we obtained
($z$ = 0.14) places this galaxy outside the planned Advanced LIGO horizon.

Finally, one object that passed our cuts, XMMSL1 J152408.6+705533,
was observed twice in the XMMSL1, in January of 2006 and November of 2007.
In the 23 months between the observations, the source faded by a 
factor of 3.  Fitting a TDE light curve to these two points 
resulted in an event with starting time in March of 2004, and a luminosity 
of 5.7$\times 10^{43}$ erg s$^{-1}$ 1 year after the start time.  The 
implied energies were roughly consistent with previously observed 
TDEs \citep{esquej08}.  We obtained a spectrum of the host galaxy, MRK 1097,
with the 200 inch telescope at Palomar observatory, and found primarily 
narrow line emission
features (see Figure \ref{spec_J1524}).  The WISE colors for this 
galaxy ([$W1-W2$] = 0.3 mag and [$W2-W3$] = 1.8 mag) \citep{wright} and narrow line features suggested
this is a spiral galaxy, and so is unlikely to host an AGN. 

\subsection{GRB Afterglows}

Short GRBs have been observed with a rate density 
of 5 to 13 Gpc$^{-3}$yr$^{-1}$, implying a rate of around
1 per year within 300 Mpc \citep{nakar07, coward2012}.  Long GRBs are observed
with a rate density of 0.5 Gpc$^{-3}$yr$^{-1}$ \citep{nakar07}, or
one event every 20 years within 300 Mpc.  GRBs are known
to display bright afterglows, typically observable in 
soft X-rays for a few hours up to a few days after
the burst.  However, most GRBs are observed much
further away than 300 Mpc, so a burst at the 
distance of our objects would have an afterglow
that would appear brighter, and would potentially
be observable longer.  Taking an optimistic but plausible
scenario, a short GRB afterglow at the distances our objects, 
showing power-law dimming with a temporal index of 1.2,
would display the flux levels observed with XMM-Newton
10 to 100 days after the burst.  A model that includes
a jet break would show a faster fading, with a temporal
index closer to 2.0 \citep{judy09}.  This means that, even in the optimistic
case, we might only expect one GRB per year within
our search volume, and that a GRB's afterglow would only be observable
for one to three months.  Based on these numbers,
we believe there is less than a $\sim$10\% chance that our candidate
list, after the requirement of a galaxy association,
includes one or more GRB afterglows.  

On the other hand, there are related classes of transients,
both observed and theoretical, that are thought to be more common
in the local universe.  Low-luminosity GRBs have been observed
with a local rate density much higher than the rate of cosmologically observed
bright GRBs \citep{soderberg2006, cobb2006, chapman2007}.  
These events can have X-ray band afterglows that are less 
luminous than cosmological GRB afterglows, but otherwise with similar 
properties \citep{soderberg2006}.  
So-called orphan afterglows and failed GRBs may also be more common in the local
universe than the more commonly observed cosmological 
GRBs \citep{rhoads2003, huang2002},
an idea that may be supported by a recent observation \citep{cenko2013}.
It is difficult to rule out the possibility that such an event is 
in our sample.

\subsection{Ultra-luminous X-ray Source}

Ultra-luminous X-ray Sources (ULXs) have been observed in several
nearby galaxies, with X-ray luminosities of $10^{38} - 10^{41}$
erg s$^{-1}$.  These objects are known to exhibit short time variability,
and have been observed in both high and low energy states \citep{winter}.  However,
the luminosities of our sources exceed the range of known ULXs.  
So, these objects cannot be naturally characterized as 
ULXs, or at best, they would represent extreme examples of the class.

\subsection{AGN}

We inspected optical spectra 
for five of our six objects, none of which showed 
evidence for an active central region.  Two of the 
spectra showed no strong emission lines, while two
host galaxies (SDSS J084838.57+193528.9 and MRK 1097) 
showed star-formation lines.  
It is possible that one or both of the galaxies for which we do 
not have optical spectra will turn out to be AGN.
Moreover, there are known cases of galaxies that seem 
consistent with an AGN model when observed in X-rays,
but do not show evidence of nuclear activity in their 
optical spectra \citep{jackson12}.  The details of the mechanism that 
hides the active region is still being disputed.  
While AGN are known to exhibit variability, both on long and 
short time-scales, the majority of this variability
is low amplitude.  For example, \citet{saxton11} found that,
in a sample of over a 1000 AGN observed with both the 
XMM-Newton Slew Survey and RASS, only 5\% varied by more than a factor 
of ten.  Given the relative rarity of large amplitude variability
in AGNs, and the current difficulty in describing optically 
quiescent but X-ray luminous galaxies, a low redshift, variable, ``hidden''
AGN might be an interesting source for future study.  

\subsection{Spurious Detections} \label{spurious}

Given that we have selected objects that appeared in the XMMSL1, 
but not in RASS,
one has to consider the possibility that these are spurious detections, or
perhaps that they are real sources, but the galactic associations are
incorrect.  We note that the XMMSL1 is estimated to contain less than
1\% false sources in the soft band \citep{xmm_slew}, 
though the stronger argument
is that the associations with host
galaxies are unlikely to all be false.  To make a conservative estimate of the
chance of false coincidences, we assumed a 30 arcsecond match radius, and
imagined spreading the 94 objects in our sample with flux ratio greater than
ten randomly across the sky, so that there were $1.7\times 10^{-3}$ transient
objects per deg$^2$.  We found that the odds of finding the observed number of
coincidences by chance were very small.  For example, within 200 Mpc there are
$\sim 150000$ known galaxies (3.6 galaxies per deg$^2$) \citep{mansi_phd}.  So,
within 200 Mpc, our search had a 7\% chance of finding even a single
coincidence by chance, where we found four.  Moreover, all but one of the sources matched
their host galaxies within 15" despite a 30" search radius, and none of the
transients were spatially inconsistent with having originated from within their
hosts' angular extent (see Table 2).  For these reasons, most or all of the claimed galaxy
associations are very likely real.  It is worth noting, however,
that one object has a slightly larger offset from the host galaxy, namely
J202320.7-670021.  The matched host galaxy is at a distance of only 
67 Mpc, and the angular extent of the galaxy overlaps the 1-sigma error
circle from XMMSL1.  So, while it seems likely that this association is real,
in this case we are unable to rule out a spurious association.  

\section{Conclusions}

In this work, we took a census of X-ray transient objects in the 
low redshift universe, motivated by future observations of compact object 
mergers with Advanced LIGO and Advanced Virgo.
A wide-field, soft X-ray monitor such as the proposed ISS-Lobster
will be able to seek counterparts to LIGO/Virgo events, but high-confidence
identifications will demand high quality variability studies.
We performed a systematic search for 
low-redshift, extragalactic transients in the XMM-Newton slew survey, 
covering 32,800 square degrees, above fluxes of 3$\times10^{-12}$ erg cm$^{-2}$ s$^{-1}$ 
in the 0.2-2 keV band.
We compared observations taken many years apart in the RASS and the 
XMMSL1, and found that over these long time scales, variations 
in flux divided sources into two categories: continuum variability, 
which can be described as a log-normal distribution in flux variability
with a width of around 3, and state-change variability, corresponding to
objects which show more dramatic changes in flux. 
State-change sources at low red-shift may be confusion sources
for future searches for counterparts to events measured with 
Advanced LIGO and Advanced Virgo, so we sought to characterize their 
density on the sky.
We found that transient sources represented around 10\%
of all objects in a flux limited survey, and that of these, 
around 10\% could be associated with known optical galaxies.
For searches for LIGO/Virgo counterparts using wide-field
X-ray imagers capable of observing hundreds of square degrees,
we should expect around one false coincidence for every 10,000 
square degrees searched to a flux limit of 10$^{-11}$ erg s$^{-1}$ cm$^{-2}$.

We found twelve objects meeting our search criteria, 
most of which were located within 350 Mpc.
Of these, we identified six with clear galaxy identifications 
that were difficult to characterize
and that have not been previously studied.  They were highly luminous, 
highly variable, and lacked classical AGN
optical signatures in the five cases with available spectra.   
Four of the sources (XMMSL1 J131951.9+225957, J084837.9+193537, J182609.9+545005, and J152408.6+705533)
met all of the criteria of \citet{esquej07} for identifying candidate
tidal disruption events.  One other seemed consistent with a TDE description,
but may prove to be a variable AGN, and the sixth source had a measured position only 
marginally consistent with the center of the apparent host galaxy.
It is possible that, whatever the nature
of these sources, they represent some of the closest members of the 
class of unexplained, extragalactic transients identified by \citet{starling11},
who focused on transient objects \textit{not} associated with known galaxies.  
This study represents the first attempt to characterize soft X-ray 
confusion sources for counterparts to Advanced LIGO/Virgo merger events, 
and in the process, has revealed a handful of unusual objects that 
are both powerful and dynamic.

\acknowledgments

The authors are grateful for helpful discussions with Suvi Gezari,
Judith Racusin, Brennan Hughey, Tracy Huard, and Sjoert van Velzan.  
We thank the anonymous referee for helpful comments.  
JK and LB were supported by appointments to the NASA Postdoctoral Program at GSFC, 
administered by Oak Ridge Associated Universities through a contract with NASA.
K.M. would like to thank Branimir Sesar, Eric Bellm, and Yi Cao for 
help obtaining the P200 spectra, and Assaf Horesh for work with the
Keck Observatory.
  
This research has made use of the NASA/IPAC Extragalactic Database (NED) which is 
operated by the Jet Propulsion Laboratory, California Institute of Technology, under contract with the National Aeronautics and Space Administration.
This research has made use of data obtained from the High Energy Astrophysics Science Archive Research Center (HEASARC), provided by NASA's Goddard Space Flight Center.  

    Funding for the SDSS and SDSS-II has been provided by the Alfred P. Sloan Foundation, the Participating Institutions, the National Science Foundation, the U.S. Department of Energy, the National Aeronautics and Space Administration, the Japanese Monbukagakusho, the Max Planck Society, and the Higher Education Funding Council for England. The SDSS Web Site is http://www.sdss.org/.

    The SDSS is managed by the Astrophysical Research Consortium for the Participating Institutions. The Participating Institutions are the American Museum of Natural History, Astrophysical Institute Potsdam, University of Basel, University of Cambridge, Case Western Reserve University, University of Chicago, Drexel University, Fermilab, the Institute for Advanced Study, the Japan Participation Group, Johns Hopkins University, the Joint Institute for Nuclear Astrophysics, the Kavli Institute for Particle Astrophysics and Cosmology, the Korean Scientist Group, the Chinese Academy of Sciences (LAMOST), Los Alamos National Laboratory, the Max-Planck-Institute for Astronomy (MPIA), the Max-Planck-Institute for Astrophysics (MPA), New Mexico State University, Ohio State University, University of Pittsburgh, University of Portsmouth, Princeton University, the United States Naval Observatory, and the University of Washington.

\bibliographystyle{apj}
\bibliography{references}
\end{document}